\documentclass{rspublic}
\usepackage{natbib}
\usepackage{graphicx}
\usepackage{color}
\usepackage{textcomp}
\usepackage{graphics} 
\usepackage{epsfig}
\usepackage{psfrag}
\usepackage{amsmath}
\usepackage{amssymb}
\usepackage{amsthm}
\usepackage{amsfonts}
\usepackage{times}
\usepackage[outerbars,color]{changebar}
\usepackage{mathptmx} 
\usepackage[all]{xy}

\cbcolor{red}

\def\supit#1{\raisebox{0.8ex}[8pt]{\footnotesize #1}\hspace{0.05em}}

\begin{document}

\title[The Fourth Element]{The Fourth Element: Characteristics, Modelling, and Electromagnetic Theory of the Memristor}

\author[O.\ Kavehei et.\ al.]{O.\ Kavehei\supit{1}, A.\ Iqbal\supit{1}, Y.\ S.\ Kim\supit{1,2}, K.\ Eshraghian\supit{3}, S.\ F.\ Al-Sarawi\supit{1}, and D.\ Abbott\supit{1}}

\affiliation{ \supit{1} School of Electrical and Electronic Engineering, University of Adelaide, \\ Adelaide, SA 5005, Australia \\ \supit{2} Department of Semiconductor Engineering, Chungbuk National University, \\ Cheongju, South Korea \\ \supit{3} College of Elec. and Info. Engineering, WCU Program, Chungbuk National University, \\ Cheongju, South Korea}

\label{firstpage}

\maketitle

\begin{abstract}{Memristor, SPICE macro-model, Nonlinear circuit theory, Nonvolatile memory, Dynamic systems}
In 2008, researchers at HP Labs published a paper in {\it Nature} reporting the realisation of a new basic circuit element that completes the missing link between charge and flux-linkage, which was postulated by Leon Chua in 1971. The HP memristor is based on a nanometer scale TiO$_2$ thin-film, containing a doped region and an undoped region. Further to proposed applications of memristors in artificial biological systems and nonvolatile RAM (NVRAM), they also enable reconfigurable nanoelectronics. Moreover, memristors provide new paradigms in application specific integrated circuits (ASICs) and field programmable gate arrays (FPGAs). A significant reduction in area with an unprecedented memory capacity and device density are the potential advantages of memristors for Integrated Circuits (ICs). This work reviews the memristor and provides mathematical and SPICE models for memristors. Insight into the memristor device is given via recalling the quasi-static expansion of Maxwell's equations. We also review Chua's arguments based on electromagnetic theory.

\end{abstract}

\section{Introduction} \label{intro}	Based on the {\it International Technology Roadmap for Semiconductors} (ITRS) report~\citep{c18}, it is predicted that by 2019, $16~{\rm nm}$ half-pitch Dynamic Random Access Memory (DRAM) cells will provide a capacity around $46~{\rm GB}/{\rm cm^2}$, assuming $100\%$ area efficiency. Interestingly, memristors promise extremely high capacity more than $110~{\rm GB}/{\rm cm^2}$ and $460~{\rm GB}/{\rm cm^2}$ for $10~{\rm nm}$ and $5~{\rm nm}$ half-pitch devices, respectively~\citep{c13, 3DIC_2009}. In contrast to DRAM memory, memristors provide nonvolatile operation as is the case for flash memories. Hence, such devices can continue the legacy of Moore's law for another decade. Furthermore, inclusion of molecular electronics and computing as an alternative to CMOS technologies in the recent ITRS report emphasises the significant challenges of device scaling~\citep{Computing_with_molecules}. Moreover,~\citet{Programmable_current_mode_Hebbian} demonstrated that the complexity of a synapse, in an analog VLSI neural network implementation, is minimised by using a device called the  {\it Programmable Metallization Cell} (PMC). This is an ionic programmable resistive device and a memristor can be employed to play the same role.
	
	Research on memristor applications in various areas of circuit design, alternative materials and spintronic memristors, and especially memristor device/circuit modeling have appeared in the recent literature, (i) SPICE macro-modeling using linear and non-linear drift models~\citep{Compact_modeling_and_corner_analysis_of_spintronic_memristor, ICCCAS09_Approximated_SPICE_Model_for_Memristor, SPICE_Model_of_memristor_with_nonlinear_dopant_drift, On_SPICE_macromodelling_of_Ti, ICCCAS09_our_paper, ICCCAS09_A_PWL_Model_of_memristor}, (ii) Application of memristors in different circuit configurations and their dynamic behaviour~\citep{ICCCAS09_Chaos_generator_based_on_a_PWL_memristor, ICCCAS09_chaotic_oscillator, ICCCAS09_Chaos_in_memristor, ICCCAS09_A_simple_memristor_chaotic_oscillator}, (iii) Application of memristor based dynamic systems to image encryption in~\citet{ICCCAS09_Image_Encryption}, (iv) Fine resolution programmable resistor using a memristor in~\citet{ICCCAS09_memristor-based_fine_resolution}, (v) Memristor-based op-amp circuit and filter characteristics of memristors by~\citet{ICCCAS09_Op-Amp} and~\citet{ICCCAS09_Study_of_filter_characteristics_based_on_PWL_memristor}, respectively, (vi) Memristor receiver (MRX) structure for ultra-wide band (UWB) wireless systems~\citep{Memristor_Oscillators, Memristor_based_stored_reference_receiver_the_UWB_solution}, (vii) Memristive system I/O nonlinearity cancellation in~\citet{Input_Output_linearization_of_memristive_systems}, (viii) The number of required memristors to compute a $f:~\mathbb{R}^n ~\rightarrow ~\mathbb{R}^m$ function by~\citet{Stateful_implication_logic_with_memristors}, and its digital logic implementation using a memristor-based crossbar architecture in~\citet{ICCCAS09_Digital_Logic}, (ix) Different physical mechanisms to store information in memristors~\citep{Phase_transition_driven_memristive_system, Spintronic_Memristor_Through_Spin-Torque_Induced_Magnetization_Motion}, (x) Interesting fabricated nonvolatile memory using a~{\it flexible memristor}, which is an inexpensive and low-power device solution is also recently reported~\citep{A_Flexible_Solution_Processed_Memristor}, (xi) Using the {\it memductance} concept to develop an equivalent circuit diagram of a transmission-line has been carried out by~\citet[appendix~2]{PHDTHESIS_French}.

	There are still many problems associated with memristor device level. For instance, it is not clear that which leakage mechanisms are associated with the device. The {\it flexible memristor} is able to retain its nonvolatile characteristic for up to 14 days or, equivalently, up to 4000 flexes~\citep{A_Flexible_Solution_Processed_Memristor}. Thus, the nonvolatility feature eventually vanishes after a short period.~\citet{Exponential_ionic_drift_fast_switching_and_low_volatility_of_thin-film_memristors} also investigated this particular feature as a ratio of volatility to switching time.

One disadvantage of using memristors is switching speed. The volatility-to-switching speed ratio for memristor cells, in the HP cross-bar structure is around $10^3$~\citep{Exponential_ionic_drift_fast_switching_and_low_volatility_of_thin-film_memristors}, which is much lower than the ratio for DRAM cells, $10^6$~\citep{c18}, therefore, the switching speed of memristors is far behind DRAM. However, unlike DRAM, RRAM is non-volatile. In terms of yield, DRAM and RRAM are almost equal~\citep{3DIC_2009}. Generally, endurance becomes very important once we note that the DRAM cells must be refreshed at least every $16~{\rm ms}$, which means at least $10^{10}$ write cycles in their life time~\citep{3DIC_2009}. Unfortunately, memristors are far behind DRAM in terms of the endurance view point, but the HP team is confident that there is no functional limitation against improvement of memristor~\citep{c13}. Finally, another advantage of RRAM is readability. Readability refers to the ability of a memory cell to report its state. This noise immunity in DRAM is weak because each DRAM cell stores a very small amount of charge particularly at nanometer dimensions, while in RRAM cells, e.g. memristors, it depends on the difference between the on and off state resistances~\citep{3DIC_2009}. This difference was reported up to one order of magnitude~\citep{c13}.

It is interesting to note that there are devices with similar behaviour to a memristor, e.g.~\citep{ Binary_information_storage_at_zero_bias_in_quantum-well_diodes, Reproducible_switching_effect_IBM, Molecular_analogue_memory_cell,
Electric-pulse-induced,
Nanoionics_based_resistive_switching_memories, Resistance_switching_in_oxide_thin_films, Reproducible_resistive_switching_effect}, but the HP scientists were the only group that found the link between their work and the missing memristor postulated by Chua. Moreover, it should be noted that physically realised memristors must meet the mathematical requirements of a memristor device or memristive systems that are discussed in~\citet{On_the_modeling_of_some_classes_of_nonlinear_devices_and_systems} and~\citet{c23}. For instance, the hysteresis loop must have a double-valued bow-tie trajectory. However, for example in~\citet{Molecular_analogue_memory_cell}, the hysteresis loop shows more than two values for some applied voltage values.~\citet{Reproducible_switching_effect_IBM} demonstrates one of the perfect examples of memristor devices.~\citet{Multilevel_Programmable_Oxide_Diode} recently introduced a multilevel one-time programmable (OTP) oxide diode for crossbar memories. They focused on a one-time programmable structure that basically utilises one diode and one resistor, 1D-1R, since obtaining a stable device for handling multiple programming and erasing processes is much more difficult than a one-time programmable device. In terms of functionality, OTP devices are very similar to memristive elements, but in terms of flexibility, memristors are able to handle multiple programming and erasing processes.

	This paper focuses on the memristor device and reviews its device level properties. Although the memristor as a device is new, it was conceptually postulated by~\citet{c16}. Chua predicted that a memristor could be realised as a purely dissipative device as a fourth fundamental circuit element. Thirty seven years later, Stan Williams and his group in the Information and Quantum Systems (IQS) Lab at HP realised the memristor in device form \citep{c17}.
	
	In Section~\ref{memristor}, we review the memristor and its characteristics as a nano-switch, which was realised by Hewlett-Packard (HP)~\citep{c17}, and we review its properties based on the early mathematical models. We introduce a new model using a parameter we call the {\it resistor modulation index} (RMI). Due to the significance of ionic drift that plays the most important role in the memristive effect, this section is divided into two: (\ref{lineardriftmodelSection}) a linear, and (\ref{nonlineardriftmodelSection}) a nonlinear drift model. Section~\ref{spice} presents a preliminary SPICE macro-model of the memristor and different types of circuit elements in combination with a proposed memristor macro-model. Section~\ref{maxwell} describes an interpretation of the memristor based on electromagnetic theory by recalling Maxwell's equations. Finally, we summarise this review in Section~\ref{conclusion}.
	

\section{Memristor Device Properties} \label{memristor}	Traditionally there are only three fundamental passive circuit elements: capacitors, resistors, and inductors, discovered in 1745, 1827, and 1831, respectively. However, one can set up five different mathematical relations between the four fundamental circuit variables: electric current $i$, voltage $v$, electric charge $q$, and magnetic flux $\varphi$. For {\it linear} elements, $f(v,i)=0$, $f(v,q)=0$ where $i=\frac{dq}{dt}$ ($q=Cv$), and  $f(i,\varphi )=0$ where $v=\frac{d\varphi }{dt}$ ($\varphi =Li$), indicate linear resistors, capacitors, and inductors, respectively.

In 1971, Leon Chua, proposed that there should be a fourth fundamental passive circuit element to set up a mathematical relationship between $q$ and $\varphi$, which he named the \textit{memristor} (a portmanteau of \textit{memory} and \textit{resistor}) \citep{c16}. Chua predicted that a class of memristors might be realisable in the form of a pure solid-state device without an internal power supply.
	
	In 2008, Williams et al., at Hewlett Packard, announced the first fabricated memristor device \citep{c17}. However, a resistor with memory is not a new thing. If taking the example of nonvolatile memory, it dates back to 1960 when Bernard Widrow introduced a new circuit element named the \textit{memistor}~\citep{c19}. The reason for choosing the name of memistor is exactly the same as the memristor, a resistor with memory. The memistor has three terminals and its resistance is controlled by the time integral of a control current signal. This means that the resistance of the memistor is controlled by charge. Widrow devised the memistor as an electrolytic memory element to form a basic structure for a neural circuit architecture called ADALINE (ADAptive LInear NEuron), which was introduced by him and his postgraduate student, Marcian Edward ``Ted'' Hoff~\citep{c19}. However, the memistor is not exactly what researchers were seeking at the nanoscale. It is just a charge-controlled three-terminal (transistor) device. In addition, a two-terminal nano-device can be fabricated without nanoscale alignment, which is an advantage over three-terminal nano-devices~\citep{Stateful_implication_logic_with_memristors}. Furthermore, the electrochemical memistors could not meet the requirement for the emerging trend of solid-state integrated circuitry.

	Thirty years later,~\citet{Solid-state_thin-film_memistor} introduced a solid-state thin-film tungsten-oxide-based, nonvolatile memory. The concept is almost similar to the HP memristors. Their solid-state memistor is electrically reprogrammable, it has variable resistance, and it is an analog synaptic memory connection that can be used in electrical neural networks. They claimed that the resistance of the device could be stabilised at any value, between $100~{\rm k\Omega}$ and $1~{\rm G\Omega}$. This solid-state memistor has a thick, $60$-$80~{\rm nm}$, layer of silicon dioxide that achieves nonvolatility over a period of several months. This thick electron blocking layer, however, causes a considerable reduction in the applied electric field. As a consequence, they reported very high programming voltages ($\pm$~25 to 30 ) and very long (minutes to hours) switching times.

	In the 1960s, the very first report on the hysteresis behaviour of current-voltage curve was published by~\citet{simmons_tunneling:1793}, which is known as the~\emph{Simmons tunneling theory}. The Simmons theory generally characterises the tunneling current in a Metal-Insulator-Metal (MIM) system. A variable resistance with hysteresis was also published by~\citet{New_Conduction_and_Reversible_Memory_Phenomena}. They introduced a thin-film ($20$ to $300~{\rm nm}$) silicon dioxide doped with gold ions sandwiched between two $200~{\rm nm}$ metal contacts. Thus overall the device is a MIM system, using aluminum metal contacts. Interestingly, their system demonstrates a sinh function behaviour as also recently reported in~\citet{c26} for the HP memristor. It is also reported that the switching from high- to low impedance takes about $100~{\rm ns}$. As their device is modelled as an energy storage element it cannot be a memristor, because memristors remember the total charge passing through the port and do not store charge~\citep{c24}.
	
	The sinh resistance behaviour of memristors can be utilised to compensate the linearity of analog circuits. In~\citet{ICCCAS09_Cadence} a memristor based amplifier was proposed utilising the behaviour of a sinh resistor~\citep{A_sinh_resistor} as a memristor element. In~\citet{ICCCAS09_memristor-based_fine_resolution} such a structure was also introduced as an example without mentioning the sinh resistance behaviour of memristors. The idea in~\citet{A_sinh_resistor} is to characterise a sinh function type circuit that can be used to linearise a tanh function type circuit behaviour, e.g. CMOS differential amplifier. A theoretical analysis shows that a sinh function significantly reduces the third harmonic coefficient and as a consequence reduces nonlinearity of circuit.

\begin{figure}[thpb]
		\centering
		\includegraphics[scale=0.07]{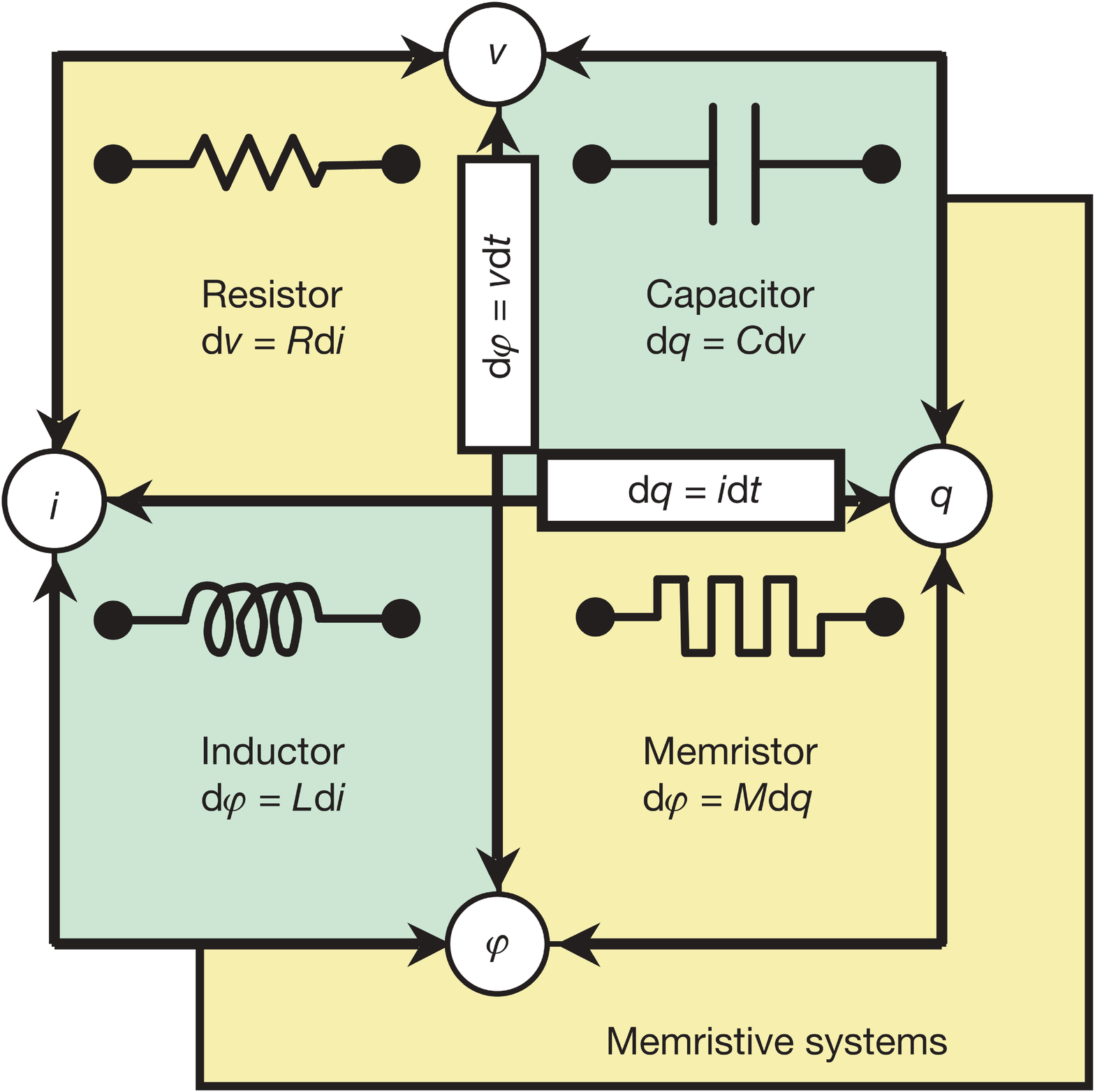}
		\caption{The four fundamental two-terminal circuit elements. There are six possible relationships between the four
fundamental circuit variables, current ($i$), voltage ($v$), charge ($q$), and magnetic flux ($\varphi$). Out of these, five relationships are known comprising, $i=\frac{dq}{dt}$ and $v=\frac{d\varphi}{dt}$, and three circuit elements: resistor (R), capacitor (C), and inductor (L). Resistance is the rate of voltage change with current, inductance is the rate of flux change with current, capacitance is the rate of charge change with voltage. The only missing relationship is the rate of flux change with voltage (or vice-versa) that leads to the fourth element, a memristor. Adapted from~\citet{c17}.}
		\label{fig6}
	\end{figure}

	Chua mathematically predicted that there is a solid-state device, which defines the missing relationship between four basic variables \citep{c16}. Recall that a resistor establishes a relation between voltage and current, a capacitor establishes a charge-voltage relation, and an inductor realises a current-flux relationship, as illustrated in Fig.~\ref{fig6}. Notice that we are specifically discussing nonlinear circuit elements here.

	Consequently, $\varphi={f_{\rm M}(q)}$ or $q={g_{\rm M}(\varphi)}$ defines a charge-controlled (flux-controlled) memristor. Then, $\frac{d\varphi}{dt}=\frac{df_{\rm M}(q)}{dq}\frac{dq}{dt}$ or $\frac{dq}{dt}=\frac{dg_{\rm M}(\varphi)}{d\varphi}\frac{d\varphi}{dt}$, which implies $v(t)=\frac{df_{\rm M}(q)}{dq}i(t)$ or $i(t)=\frac{dg_{\rm M}(\varphi)}{d\varphi}v(t)$. Note that, $M(q)=\frac{df_{\rm M}(q)}{dq}$ for a charge-controlled memristor and $W(\varphi)=\frac{dg_{\rm M}(\varphi)}{d\varphi}$ for a flux-controlled memristor, where $M(q)$ is the incremental memristance and $W(\varphi)$ is the incremental memductance where $M(q)$ is in the units of Ohms and the units of $W(\varphi)$ is in Siemens~\citep{c16}.
	
	Memristance, $M(q)$, is the slope of the $\varphi$-$q$ curve, which for a simple case, is a piecewise $\varphi$-$q$ curve with two different slopes. Thus, there are two different values for $M(q)$, which is exactly what is needed in binary logic. For detailed information regarding typical $\varphi$-$q$ curves, the reader is referred to~\citet{c16}.

	It is also obvious that if $M(q)\geq0$, then instantaneous power dissipated by a memristor, $p(i)=M(q)(i(t))^2$, is always positive so the memristor is a passive device. Thus the $\varphi$-$q$ curve is a monotonically-increasing function. This feature is exactly what is observed in the HP memristor device~\citep{c17}. Some other properties of the memristor such as zero-crossing between current and voltage signals can be found in~\citet{c16} and~\citet{c23}. The most important feature of a memristor is its pinched hysteresis loop $v$-$i$ characteristic~\citep{c23}. A very simple consequence of this property and $M(q)\geq0$ is that such a device is purely dissipative like a resistor.

	Another important property of a memristor is its excitation frequency. It has been shown that the pinched hysteresis loop is shrunk by increasing the excitation frequency~\citep{c23}. In fact, when the frequency increases toward infinity, the memristor acts as a linear resistor~\citep{c23}.

\begin{figure}[thpb]
		\centering
		\includegraphics[scale=.4]{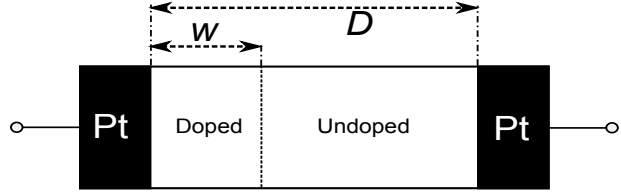}
		\caption{Schematic of HP memristor, where $D$ is the device channel length and $w$ is the length of doped region. The size of doped region is a function of the applied charge and is responsible for memristive effect as it changes the effective resistance of the device. Usually $w$ is shown by its normalised counterpart, $x=w/D$. Adapted from~\citet{c17}.}
		\label{fig8}
	\end{figure}

	Interestingly enough, an attractive property of the HP memristor \citep{c17}, which is exclusively based on its fabrication process, can be deduced from the HP memristor simple mathematical model \citep{c17} and is given by,

	\begin{equation}
		M(q)=R_{\rm OFF}\Bigg(1-\frac{R_{\rm ON}}{\beta}q(t)\Bigg)~,
		\label{equ1}
	\end{equation}
where $\beta$ has the dimensions of magnetic flux $\varphi(t)$. Here, $\beta=\frac{D^2}{\mu_{\rm D}}$ in units of ${\rm sV} \equiv {\rm Wb}$, where $\mu_{\rm D}$ is the average drift mobility in unit of ${\rm cm}^2/{\rm s V}$ and $D$ is the film (titanium dioxide, TiO$_{\rm 2}$) thickness. Note that $R_{\rm OFF}$ and $R_{\rm ON}$ are simply the `on' and `off' state resistances as indicated in Fig.~\ref{fig8}. Also $q(t)$ defines the total charge passing through the memristor device in a time window, $t$ - $t_0$. Notice that the memristor has an internal state~\citep{c23}. Furthermore, as stated in~\citet{c24}, $q(t)=\int_{t_0}^{t}i(\tau)d\tau$, as the state variable in a charge-controlled memristor gives the charge passing through the device and does not behave as storage charge as in a capacitor as incorrectly reported in some works, e.g.~\citet{New_Conduction_and_Reversible_Memory_Phenomena}. This concept is very important from two points of view. First of all, a memristor is not an energy-storage element. Second, this shows that the memristor is not merely a nonlinear resistor, but is a nonlinear resistor with charge as a state variable~\citep{c24}.
	
	Five years after Chua's paper on the memristor \citep{c16}, he and his graduate student, Kang, published a paper defining a much broader class of systems, named \textit{memristive systems}. From the memristive systems viewpoint a generalized definition of a memristor is $v(t)=R(w)i(t)$, where $w$ defines the internal state of the system and $\frac{dw}{dt}=f(w,i)$~\citep{c23}. Based on this definition a memristor is a special case of a memristive system.
	
	The HP memristor~\citep{c17} can be defined in terms of memristive systems. It exploits a very thin-film TiO$_{\rm 2}$ sandwiched between two platinum (Pt) contacts and one side of the TiO$_{\rm 2}$ is doped with oxygen vacancies, which are positively charged ions. Therefore, there is a TiO$_{\rm 2}$ junction where one side is doped and the other is undoped. Such a doping process results in two different resistances: one is a high resistance (undoped) and the other is a low resistance (doped). Hence, HP intentionally established a device that is illustrated in Fig.~\ref{fig8}. The internal state variable, $w$, is also the variable length of the doped region. Roughly speaking, when $w\rightarrow 0$ we have nonconductive channel and when $w\rightarrow D$ we have conductive channel. The HP memristor switching mechanism is further discussed in~\citet{c26}.

\subsection{Linear drift model}
\label{lineardriftmodelSection}	

	The memristor's state equation is at the heart of the HP memristive system mechanism~\citep{c25, The_elusive_memristor}. Let us assume a uniform electric field across the device, thus there is a linear relationship between drift-diffusion velocity and the net electric field~\citep{Electrocoloration_in_SrTiO}. Therefore the state equation is,

	\begin{equation}
		\frac{1}{D}\frac{dw(t)}{dt}=\frac{R_{\rm ON}}{\beta}i(t)~.
		\label{equ2}
	\end{equation}
	
	Integrating Eq.~\ref{equ2} gives $\frac{w(t)}{D}=\frac{w(t_{\rm 0})}{D}+\frac{R_{\rm ON}}{\beta}q(t)$, where $w(t_{\rm 0})$ is the initial length of $w$. Hence, the speed of drift under a uniform electric field across the device is given by $v_{\rm D}=\frac{dw(t)}{dt}$. In a uniform field we have $D=v_{\rm D} \times t$. In this case $Q_{\rm D}=i \times t$ also defines the amount of required charge to move the boundary from $w(t_{\rm 0})$, where $w\rightarrow 0$, to distance $w(t_{\rm D})$, where $w\rightarrow D$. Therefore, $Q_{\rm D}=\frac{\beta}{R_{\rm ON}}$, so

	\begin{equation}
		\frac{w(t)}{D}=\frac{w(t_{\rm 0})}{D}+\frac{q(t)}{Q_{\rm D}}~.
		\label{equ9}
	\end{equation}

	If $x(t)=\frac{w(t)}{D}$ then

	\begin{equation}
		x(t)=x(t_{\rm 0})+\frac{q(t)}{Q_{\rm D}}~,
		\label{equ9.1}
	\end{equation}
where $\frac{q(t)}{Q_{\rm D}}$ describes the amount of charge that is passed through the channel over the required charge for a conductive channel.

Using~\citet{c17} we have,

	\begin{equation}
		v(t)=\Bigg(R_{\rm ON}~\frac{w(t)}{D}+R_{\rm OFF} \Bigg(1-\frac{w(t)}{D}\Bigg) \Bigg)i(t)~.
		\label{equ3.1}
	\end{equation}

By inserting $x(t)=\frac{w(t)}{D}$, Eq.~\ref{equ3.1} can be rewritten as

	\begin{equation}
		v(t)=\Bigg(R_{\rm ON}~x(t)+R_{\rm OFF}\Bigg(1-x(t)\Bigg)\Bigg)i(t)~.
		\label{equ3}
	\end{equation}

Now assume that $q(t_0)=0$ then $w(t)=w(t_0)\neq 0$, and from Eq.~\ref{equ3},

	\begin{equation}
		M_0=R_{\rm ON}\Bigg(x(t_0)+r\Bigg(1-x(t_0)\Bigg)\Bigg)~,
		\label{equ6}
	\end{equation}
where $r=\frac{R_{\rm OFF}}{R_{\rm ON}}$ and $M_0$ is the memristance value at $t_0$. Consequently, the following equation gives the memristance at time $t$,

	\begin{equation}
		M(q)=M_0-\Delta R\Bigg(\frac{q(t)}{Q_{\rm D}}\Bigg)~,
		\label{equ7}
	\end{equation}

where $\Delta R=R_{\rm OFF}-R_{\rm ON}$. When $R_{\rm OFF}\gg R_{\rm ON}$, $M_0\approx R_{\rm OFF}$ and Eq.~\ref{equ1} can be derived from Eq.~\ref{equ7}.

Substituting Eq.~\ref{equ7} into $v(t)=M(q)i(t)$, when $i(t)=~\frac{dq(t)}{dt}$, we have,

	\begin{equation}
		v(t)=\Bigg(M_0-\Delta R\Bigg(\frac{q(t)}{Q_{\rm D}}\Bigg)\Bigg)\frac{dq(t)}{dt}~.
		\label{equ7.1}
	\end{equation}

Recalling that $M(q)=\frac{d\varphi(q)}{dq}$, the solution is

	\begin{equation}
		q(t)=\frac{Q_D M_0}{\Delta R}\Bigg(1\pm \sqrt{1-\frac{2\Delta R}{Q_D M^2_0}\varphi(t)}~\Bigg)~.
		\label{equ7.2}
	\end{equation}

Using $\Delta R\approx M_0\approx R_{\rm OFF}$, Eq.~\ref{equ7.2} becomes,

	\begin{equation}
		q(t)=Q_D\Bigg(1- \sqrt{1-\frac{2}{Q_D R_{\rm OFF}}\varphi(t)}~\Bigg)~.
		\label{equ7.3}
	\end{equation}

Consequently, using Eq.~\ref{equ9.1} if $Q_D=\frac{D^2}{\mu_D R_{\rm ON}}$, so the internal state of the memristor is

 	\begin{equation}
		x(t)=1-\Bigg(\sqrt{1-\frac{2 \mu_D}{r D^2}\varphi(t)}~\Bigg)~.
		\label{equ11}
	\end{equation}

The current-voltage relationship in this case is,

	\begin{equation}
		i(t)=\frac{v(t)}{R_{\rm OFF} \Bigg(\sqrt{1-\frac{2 \mu_D}{r D^2}\varphi(t)}~\Bigg)}~.
		\label{equ7.4}
	\end{equation}

	In Eq.~\ref{equ7.4}, the inverse square relationship between memristance and  TiO$_2$ thickness, $D$, shows that for smaller values of $D$, the memristance shows improved characteristics, and because of this reason the memristor imposes a small value of $D$.
	
	In Eqs.~\ref{equ7.2}-\ref{equ7.4}, the only term that significantly increases the role of $\varphi(t)$ is lower $Q_{\rm D}$. This shows that at the micrometer scale $\frac{1}{R_{\rm OFF}Q_{\rm D}}=\frac{1}{r \beta}=\frac{\mu_{\rm D}}{r D^2}$ is negligible and the relationship between current and voltage reduces to a resistor equation.
	
	Substituting $\beta=\frac{D^2}{\mu_D}$ that has the same units as magnetic flux into Eq.~\ref{equ7.4}, and considering $c(t)=\frac{\varphi(t)}{\beta}=\frac{\mu_D \varphi(t)}{D^2}$ as a normalised variable, we obtain

	\begin{equation}
		i(t)=\frac{v(t)}{R_{\rm OFF} \Bigg(\sqrt{1-\frac{2}{r}c(t)}~\Bigg)}~,
		\label{equ7.5}
	\end{equation}	
where $\sqrt{1-\frac{2}{r}c(t)}$ is what we call the \textit{resistance modulation index}, (RMI)~\citep{ICCCAS09_our_paper}.
	
	Due to the extremely uncertain nature of nanotechnologies, a variability-aware modeling approach should be always considered. Two well-know solutions to analyse a memristive system were investigated in~\citet{Compact_modeling_and_corner_analysis_of_spintronic_memristor}, 1) Monte-Carlo simulation for evaluating (almost) complete statistical behaviour of device, and 2) Corner analysis. Considering the trade-off between time-complexity and accuracy between these two approach as shows the importance of using a simple and reasonably accurate model, because finding the real corners is highly dependent on the model accuracy. The resistance modulation index, RMI, could be one of the device parameters in the model extraction phase, so it would help to provide a simple model with fewer parameters.
	
\begin{figure}[thpb]
		\centering
		\includegraphics[scale=.5]{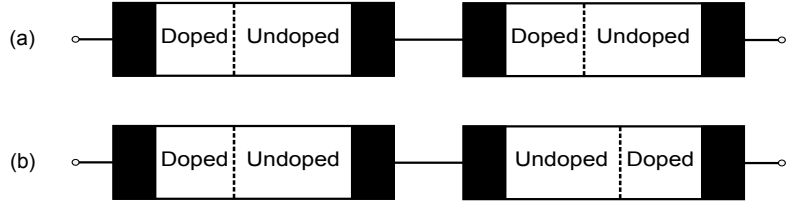}
		\caption{Two memristors in series, (a) With the same polarity, both $\eta=-1$ or both $\eta=+1$. (b) With opposite polarities, $\eta=-1$ and $\eta=+1$. Where $\eta=+1$ signifies that $w(t)$ increases with positive voltage or, in other words, the doped region in memristor is expanding and $\eta=-1$ indicates that the doped region is shrinking with applied positive voltage across the terminals. Adapted from~\citet{The_elusive_memristor}.}
		\label{polarity}
	\end{figure}

	\citet*{The_elusive_memristor} clarified the behaviour of two memristors in series. As shown in Fig.~\ref{polarity}, they labeled the polarity of a memristor by $\eta=\pm 1$, where $\eta=+1$ signifies that $w(t)$ increases with positive voltage or, in other words, the doped region in memristor is expanding. If the doped region, $w(t)$, is shrinking with positive voltage, then $\eta=-1$. In other words, reversing the voltage source terminals implies memristor polarity switching. In Fig.~\ref{polarity} (a), the doped regions are simultaneously shrunk so the overall memristive effect is retained. In Fig.~\ref{polarity} (b), however, the overall memristive effect is suppressed~\citep{The_elusive_memristor}.
	
\begin{figure}[thpb]
		\centering
		\includegraphics[scale=.5]{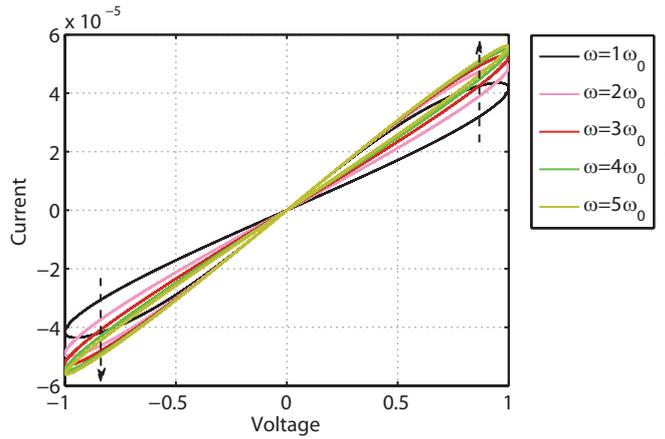}
		\caption{The hysteresis of a memristor based on Eq.~\ref{equ7.5.mod}. This verifies the hysteresis shrinks at higher frequencies. It also shows that the effective resistance is changing, so there is a varying memristance with a monotonically-increasing $q$-$\varphi$ curve.}
		\label{OurModelIV}
	\end{figure}

	Using the memristor polarity effect and Eq.~\ref{equ2}, we thus obtain
		
	\begin{equation}
		\frac{1}{D}\frac{dw(t)}{dt}=\eta \frac{R_{\rm ON}}{\beta}i(t)~.
		\label{equ2.mod}
	\end{equation}
Then with similar approach we have
	
	\begin{equation}
		i(t)=\frac{v(t)}{R_{\rm OFF} \Bigg(\sqrt{1-\eta \frac{2}{r}c(t)}~\Bigg)}~.
		\label{equ7.5.mod}
	\end{equation}
There is also no phase shift between current and voltage signals, which implies that the hysteresis loop always crosses the origin as demonstrated in Fig.~\ref{OurModelIV}. For further investigation, if a voltage, $v(t)=v_0\sin(\omega t)$, is applied across the device, the magnetic flux would be $\varphi(t)=-\frac{v_0}{\omega}\cos(\omega t)$. The inverse relation between flux and frequency shows that at very high frequencies there is only a linear resistor.

\begin{figure}[thpb]
		\centering
		\includegraphics[scale=.5]{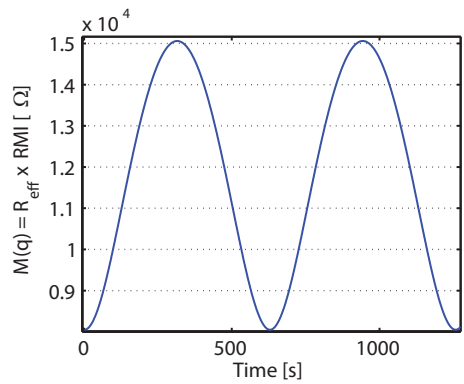}
		\caption{The hysteresis characteristics of the memristor. It shows that the memristance value is varying from a very low to a very high resistance. It is clear that these values depend on the parameter values, such as $R_{\rm on}$ and $R_{\rm off}$.}
		\label{RMI}
	\end{figure}

	Fig.~\ref{OurModelIV} demonstrates Eq.~\ref{equ7.5.mod} in MATLAB for five different frequencies, where $\omega_0 = \frac{2\pi v_0}{\beta}$, using the~\citet{c17} parameter values, $D=10~{\rm nm}$, $\mu_D=10^{-10}~{\rm cm^2~s^{-1}~V^{-1}}$, $R_{\rm ON}=100~\Omega$, $R_{\rm OFF}=16~{\rm k}\Omega$, $v_0=1~{\rm V}$, and $\eta=-1$. The \textit{resistance modulation index}, ${\rm RMI}=\sqrt{1-\eta \frac{2}{r}c(t)}$ simulation with the same parameter values is shown in Fig.~\ref{RMI}.

Using different parameters causes a large difference in the hysteresis and memristor characteristics. In~\citet{Memristor_based_stored_reference_receiver_the_UWB_solution}, a new solution for ultra-wideband signals using memristor devices was introduced. Applying the parameter values given in~\citet{Memristor_based_stored_reference_receiver_the_UWB_solution} results in a significant difference in the value of $\omega_0$. Substituting parameter values given in the paper gives $\omega_0\approx 4~{\rm GHz}$ instead of $\omega_0\approx 50~{\rm kHz}$, using to $t_0\approx 0.1~{\rm ms}$ and the parameter values from~\citet{c17}. The new parameter values are $D=3~{\rm nm}$, $\mu_D=3\times 10^{-8}~{\rm m^2~s^{-1}~V^{-1}}$, $R_{\rm ON}=100~\Omega$, $R_{\rm OFF}=10~{\rm k}\Omega$, and $v_0=0.2~{\rm V}$. Using these parameters shows that even though the highest and lowest memristance ratio in the last case, Fig.~\ref{OurModelIV}, is around $2$, here the ratio is approximately equal to $120$.

\subsection{Nonlinear drift model}
\label{nonlineardriftmodelSection}	
	
	The electrical behaviour of the memristor is directly related to the shift in the boundary between doped and undoped regions, and the effectively variable total resistance of the device when an electrical field is applied. Basically, a few volts supply voltage across a very thin-film, e.g. $10~{\rm nm}$, causing a large electric field. For instance, it could be more than $10^6~{\rm V/cm}$, which results in a fast and significant reduction in energy barrier~\citep{Electrocoloration_in_SrTiO}. Therefore, it is reasonable to expect a high nonlinearity in ionic drift-diffusion~\citep{Redox_Based_Resistive_Switching_Memories}.
	
	One significant problem with the linear drift assumption is the boundaries. The linear drift model, Eq.~\ref{equ2}, suffers from problematic boundary effects. Basically, at the nanoscale, a few volts causes a large electric field that leads to a significant nonlinearity in ionic transport~\citep{c17}. A few attempts have been carried out so far to consider this nonlinearity in the state equation~\citep{c17, Exponential_ionic_drift_fast_switching_and_low_volatility_of_thin-film_memristors, SPICE_Model_of_memristor_with_nonlinear_dopant_drift, On_SPICE_macromodelling_of_Ti}. All of them proposed a simple \textit{window function}, $F(\xi)$, which is multiplied by the right-hand side of Eq.~\ref{equ2}. In general, $\xi$ could be a variable vector, e.g. $\xi=(w,i)$ where $w$ and $i$ are the memristor's state variable and current, respectively.

\begin{figure}[thpb]
		\centering
		\includegraphics[scale=.4]{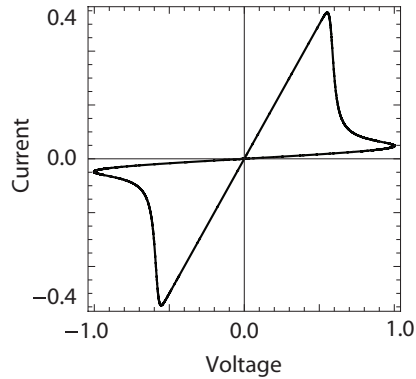}
		\caption{The hysteresis characteristics using the nonlinear drift assumption. This hysteresis shows a highly nonlinear relationship between current and voltage at the boundaries. To model this nonlinearity, there is a need for an additional term on the right hand side of the memristor's state equation, called a {\it window function}.}
		\label{linearandnonlinear}
	\end{figure}

In general, the window function can be multiplied by the right-hand side of the state variable equation, Eq.~\ref{equ2},
	
	\begin{equation}
		\frac{dx(t)}{dt}=\eta\frac{R_{\rm ON}}{\beta}i(t)F(x(t),p)~,
		\label{equ2.nonlin}
	\end{equation}
	
	where $x(t)=w(t)/D$ is the normalised form of the state variable. The window function makes a large difference between the model with linear and nonlinear drift assumptions at the boundaries. Fig.~\ref{linearandnonlinear} shows such a condition considering a nonlinear drift assumption at the critical, or boundary, states. In other words, when a linear drift model is used, simulations should consider the boundaries and all constraints on initial current, initial voltage, maximum and minimum $w$, and etc. These constraints cause a large difference in output between linear and nonlinear drift assumptions. For example, it is impossible to achieve such a realistic curve, as in Fig.~\ref{linearandnonlinear}, using the linear drift modeling approach.

In~\citet{c17}, the window function is a function of the state variable and it is defined as $F(w)=w(1-w)/D^2$. The boundary conditions in this case are $F(0)=0$ and $F(D)=\frac{1-D}{D}\approx 0$. It meets the essential boundary condition $F(\xi\rightarrow {\rm boundaries})=0$, except there is a slight difference when $w\rightarrow D$. The problem of this boundary assumption is when a memristor is driven to the terminal states, $w\rightarrow 0$ and $w\rightarrow D$, then $\frac{dw}{dt}\rightarrow 0$, so no external field can change the memristor state~\citep{SPICE_Model_of_memristor_with_nonlinear_dopant_drift}. This is a fundamental problem of the window function. The second problem of the window function is it assumes that the memristor remembers the amount of charge that passed through the device. Basically, this is a direct result of the state equation, Eq.~\ref{equ2}, ~\citep{SPICE_Model_of_memristor_with_nonlinear_dopant_drift}. However, it seems that the device remembers the position of boundary between the two regions, and not the amount of charge.

\begin{figure}[thpb]
		\centering
		\includegraphics[scale=.6]{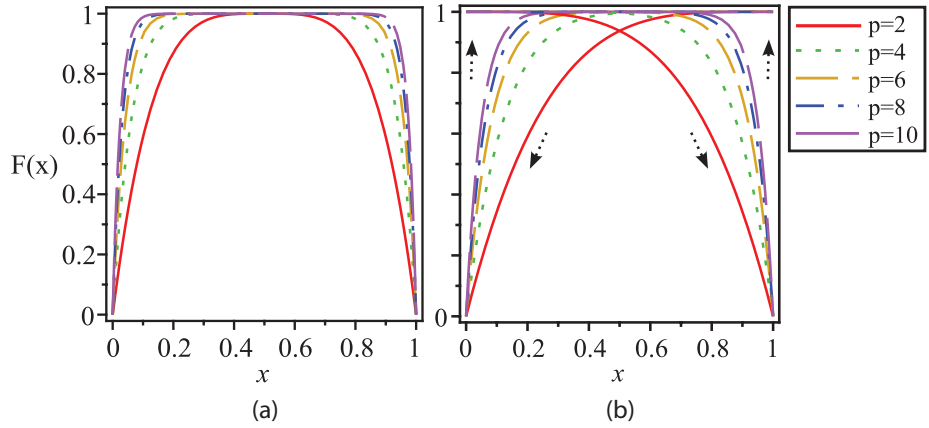}
		\caption{Non-linear window functions, (a) $F(x)=1-(2x-1)^{2p}$, (b) $F(x)=1-(x-{\rm sgn}(-i))^{2p}$. There are around four window functions in the literature but these two are meet the boundary conditions.}
		\label{nonlinear-drift}
	\end{figure}

	In~\citet{On_SPICE_macromodelling_of_Ti}, another window function has been proposed that is slightly different from that in~\citet{c17}. This window function, $F(w)=w(D-w)/D^2$, approaches zero when $w\rightarrow 0$ and when $w\rightarrow D$ then $F(w)\rightarrow 0$. Therefore, this window function meets both the boundary conditions. In fact, the second window function is imitates the first function when we consider $x=w/D$ instead of $w$. In addition, there is another problem associated with these two window functions, namely, the modeling of approximate linear ionic drift when $0<w<D$. Both of the window functions approximate nonlinear behaviour when the memristor is not in its terminal states, $w=0$ or $w=D$. This problem is addressed in~\citet{The_elusive_memristor} where they propose an interesting window function to address the nonlinear and approximately linear ionic drift behaviour at the boundaries and when $0<w<D$, respectively. Nonlinearity (or linearity) of their function can be controlled with a second parameter, which we call the \textit{control parameter}, $p$. Their window function is $F(x)=1-(2x-1)^{2p}$, where $x=w/D$ and $p$ is a positive integer.  Fig.~\ref{nonlinear-drift} (a) demonstrates the function behaviour for different $2\leq p\leq 10$ values. This model considers a simple boundary condition, $F(0)=F(1)=0$. As demonstrated, when $p\geq 4$, the state variable equation is an approximation of the linear drift assumption, $F(0<x<1)\approx 1$.

The most important problem associated with this model is revealed at the boundaries. Based on this model, when a memristor is at the terminal states, no external stimulus can change its state.~\citet{SPICE_Model_of_memristor_with_nonlinear_dopant_drift} tackles this problem with a new window function that depends on $x$, $p$, and memristor current, $i$. Basically, $x$ and $p$ are playing the same role in their model and the only new idea is using current as an extra parameter. The window function is, $F(x)=1-(x-{\rm sgn}(-i))^{2p}$, where $i$ is memristor current and ${\rm sgn}(i)=1$ when $i\geq 0$, and ${\rm sgn}(i)=0$ when $i<0$. As a matter of fact, when the current is positive, the doped region length, $w$, is expanding. Fig.~\ref{nonlinear-drift} (b) illustrates the window function behaviour.

All window functions suffer from a serious problem. They are only dependent on the state variable, $x$, which implies that the memristor remembers the entire charge that is passing through it. Moreover, based on the general definition of the time-invariant memristor's state equation and current-voltage relation, $\dot{x}=f(x,i)$ or $\dot{x}=g(x,v)$, $f$ and $g$ must be {\it continuous} $n$-dimensional vector functions~\citep[chap.~2]{On_the_modeling_of_some_classes_of_nonlinear_devices_and_systems}. However, the last window function, $F(x)=1-(x-{\rm sgn}(-i))^{2p}$, does not provide the continuity condition at the boundaries, $x\rightarrow 0$ or $x\rightarrow 1$. \citet{SPICE_Model_of_memristor_with_nonlinear_dopant_drift} did not use the window function in their recent publication~\citep{SPICE_modeling_of_memristive_memcapacitative_and_meminductive}.

In~\citet{Exponential_ionic_drift_fast_switching_and_low_volatility_of_thin-film_memristors} the overall drift velocity is identified with one linear equation and one highly nonlinear equation, $\upsilon =\mu E$, when $E<<E_0$ and $\upsilon =\mu E_0 {\rm exp}(\frac{E}{E_0})$, for $E\sim E_0$, where $\upsilon $ is the average drift velocity, $E$ is an applied electric field, $\mu$ is the mobility, and $E_0$ is the characteristic field for a particular mobile ion. The value of $E_0$ is typically about $1~{\rm MV/cm}$ at room temperature~\citep{Exponential_ionic_drift_fast_switching_and_low_volatility_of_thin-film_memristors}. This equation shows that a very high electric field is needed for exponential ion transport. They also showed that the high electric field is lower than the critical field for dielectric breakdown of the oxide. Reviewing the available window functions indicates that there is a need for an appropriate model that can define memristor states for strongly nonlinear behavior where, $E\sim E_0$.
	
	In addition to the weakness of nonlinear modeling in the original HP model, there are some other drawbacks that show the connection between physics and electronic behaviour was not well established. Moreover, the currently available electronic models for memristors followed the exact pathway of the HP modeling, which is mostly due to the fact that the underlying  physical conduction mechanism is not fully clear yet~\citep{Nanoparticle_Korean_2009}. One weakness is that the HP model does not deliver any insight about capacitive effects, which are naturally associated with memristors. These capacitive effects will later be explained in terms of a memcapacitive effect in a class of circuit elements with memory. In~\citet{Nanoparticle_Korean_2009} the memristor behaviour was realised using infinite number of crystalline magnetite (Fe$_3$O$_4$) nanoparticles. The device behaviour combines both memristive (time-varying resistance) and memcapacitive (time-varying capacitance) effects, which deliver a better model for the nonlinear properties. Their model description for current-voltage relationship is given as,

	\begin{equation}
		i(t)=\frac{v(t)}{\sqrt{R^2(x,t)+\frac{1}{i^2(t)}\Big(\frac{q(t)}{C(x,t)} \Big)^2}}~,
		\label{equ.nanopart.korea09.iv}
	\end{equation}
where $R(x,t)$ and $C(x,t)$ are the time-varying resistance and capacitance effects, respectively. The time-dependent capacitor is a function of the state variable, $x(t)=w(t)/D$ and $\Delta C(t)$, where $\Delta C(t)$ is defined as the additional capacitance caused by changing the value of capacitance~\citep{Nanoparticle_Korean_2009}, $C(x,t)=\frac{C_{\rm ON}-\Delta C(t)}{1-x(t)}$, where $C_{\rm ON}$ is the capacitance at $x=0$. The state variable is also given by,

	\begin{equation}
		x(t)=\frac{1}{\beta} \Bigg( R_{\rm ON} q(t) + \frac{\int^{t}_{t_0} \! q(\tau) \, d\tau}{C_{\rm ON}-\Delta C(t)} \Bigg)~.
		\label{equ.nanopart.korea09.x}
	\end{equation}

~\citet{Nanoparticle_Korean_2009} also investigated the impact of temperature variation on their Fe$_3$O$_4$ nanoparticle memristor assemblies of $D$ equal to $9$, $12$, and $15$ ${\rm nm}$. It was reported that the change in electrical resistivity (specific electrical resistance), $\rho_r$, as an explicit function of temperature can be defined by, ${\rm log} \rho_r = 1/ \sqrt{T}$, which means there is a significantly increasing behaviour as temperature decreases. As a consequence, for example, there is no hysteresis loop signature at room temperature, $T=295~{\rm K}$, ($D=12~{\rm nm}$) while at $T=210~{\rm K}$ it shows a nice bow-tie trajectory. As they claimed, the first room temperature reversible switching behaviour was observed in their nanoparticle memristive system~\citep{Nanoparticle_Korean_2009}.

It is worth noting that there are also two other elements with memory named the~\textit{memcapacitor} (Memcapacitive, MC, Systems) and~\textit{meminductor} (Meminductive, MI, Systems). \citet{Putting_Memory_Into_Circuit_Elements} postulated that these two elements also could be someday realised in device form. The main difference between these three elements, the memristor, memcapacitor, and meminductor is that, the memristor is not a lossless memory device and dissipates energy as heat. However, at least in theory, the memcapacitor and meminductor are lossless devices because they do not have resistance.~\citet{Putting_Memory_Into_Circuit_Elements} also investigated some examples of using different nanoparticle-based thin-film materials. Memristors (Memresistive, MR, Systems) are identified by a hysteresis current-voltage characteristic, whereas MC and MI systems introduce Lissajous curves for charge-voltage and flux-current, respectively. Similar to memristors, there are two types of these elements, therefore, the three circuit elements with memory (mem-systems/devices) can be summarised as follows,

\begin{description}

\item[Memristors] (MR Systems): A memristor is an one-port element whose instantaneous electric charge and flux-linkage, denoted by $q_{\rm mr}$ and $\varphi_{\rm mr}$, respectively, satisfy the relation $F_{\rm mr}(q_{\rm mr},\varphi_{\rm mr})=0$. It has been proven that these devices are passive elements~\citep{c17}. As discussed, they cannot store energy, so $v_{\rm mr}(t)=0$ whenever $i_{\rm mr}(t)=0$ and there is a pinched hysteresis loop between current and voltage. Thus, charge-flux curve is a monotonically-increasing function. A memristor acts as a linear resistor when frequency goes toward infinity and as a nonlinear resistor at low frequencies. Due to the nonlinear resistance effect, $\frac{dv_{\rm mr}}{dt}=R(t)\frac{di_{\rm mr}}{dt}+i_{\rm mr}(t)\frac{dR}{dt}$ should be utilised instead of $\frac{dv_{\rm mr}}{dt}=R(t)\frac{di_{\rm mr}}{dt}$. There are two types of control process,
	\begin{description}
	\item[I.] $n$th order current-controlled MR systems~\footnote{Current- (and voltage-) controlled is a better definition for memristors because they do not store any charge~\citep{c24}. In~\citet{Putting_Memory_Into_Circuit_Elements} it is specified as current- (and voltage-) controlled instead of charge- (and flux-) controlled in~\citet{c16}.}, $q_{\rm mr}=\int \! i_{\rm mr}(\tau) \, d\tau$,
		\begin{itemize}
		\item $v_{\rm mr}(t)=R(x,i_{\rm mr},t)i_{\rm mr}(t)$
		\item $\dot{x}=f_{\rm icmr}(x,i_{\rm mr},t)$
		\end{itemize}
	\item[II.] $n$th order voltage-controlled MR systems, $\varphi_{\rm mr}=\int \! v_{\rm mr}(\tau) \, d\tau$,
		\begin{itemize}
		\item $i_{\rm mr}(t)=R^{-1}(x,v_{\rm mr},t)v_{\rm mr}(t)$
		\item $\dot{x}=f_{\rm vcmr}(x,v_{\rm mr},t)$ .
		\end{itemize}
	\end{description}

\item[Memcapacitors] (MC Systems): A memcapacitor is an one-port element whose instantaneous flux-linkage and time-integral of electric charge, denoted by $\varphi_{\rm mc}$ and $\sigma_{\rm mc}$, respectively, satisfy the relation $F_{\rm mc}(\varphi_{\rm mc},\sigma_{\rm mc})=0$. The total added/removed energy to/from a voltage-controlled MC system, $U_{\rm mc}=\int \! v_{\rm mc}(\tau) \, i_{\rm mc}(\tau) \, d\tau $, is equal to the linear summation of areas between $q_{\rm mc}$-$v_{\rm mc}$ curve in the first and third quadrants with opposite signs. Due to the nonlinear capacitance effect, $\frac{dq_{\rm mc}}{dt}=i_{\rm mc}(t)=C(t)\frac{dv_{\rm mc}}{dt}+v_{\rm mc}(t)\frac{dC}{dt}$ should be utilised instead of $\frac{dq_{\rm mc}}{dt}=C(t)\frac{dv_{\rm mc}}{dt}$, so $U_{\rm mc}=\int \! v_{\rm mc}C \, dv_{\rm mc}+\int \! v_{\rm mc}^2 \, dC $. In principle a memcapacitor can be a passive, an active, and even a dissipative\footnote{Adding energy to system.} element~\citep{Putting_Memory_Into_Circuit_Elements}. If $v_{\rm mc}(t)=V_{\rm mc0}{\rm cos}(2 \pi \omega t)$ and capacitance is varying between two constant values, $C_{\rm ON}$ and $C_{\rm OFF}$, then the memcapacitor is a passive element. It is also important to note that, assuming zero charged initial state for a passive memcapacitor, the amount of removed energy cannot exceed the amount of previously added energy~\citep{Putting_Memory_Into_Circuit_Elements}. A memcapacitor acts as a linear capacitor when frequency tend to infinity and as a nonlinear capacitor at low frequencies. There are two types of control process,

	\begin{description}
	\item[I.] $n$th order voltage-controlled MC systems, $\varphi_{\rm mc}=\int \! v_{\rm mc}(\tau) \, d\tau$,
		\begin{itemize}
		\item $q_{\rm mc}(t)=C(x,v_{\rm mc},t)v_{\rm mc}(t)$
		\item $\dot{x}=f_{\rm vcmc}(x,v_{\rm mc},t)$
		\end{itemize}
	\item[II.] $n$th order charge-controlled MC systems, $\sigma_{\rm mc}=\int \! q_{\rm mc}(\tau) \, d\tau=\iint \! i_{\rm mc}(\tau) \, d\tau$,
		\begin{itemize}
		\item $v_{\rm mc}(t)=C^{-1}(x,q_{\rm mc},t)q_{\rm mc}(t)$
		\item $\dot{x}=f_{\rm qcmc}(x,q_{\rm mc},t)$ .
		\end{itemize}
	\end{description}

\item[Meminductors] (ML Systems): A meminductor is a one-port element whose instantaneous electric charge and time-integral of flux-linkage, denoted by $q_{\rm ml}$ and $\varrho_{\rm ml}$, respectively, satisfy the relation $F_{\rm ml}(\varrho_{\rm ml},q_{\rm ml})=0$. In the total stored energy equation in a ML system, $U_{\rm ml}=\int \! v_{\rm ml}(\tau) \, i_{\rm ml}(\tau) \, d\tau $, the nonlinear inductive effect, $\frac{d\varphi_{\rm ml}}{dt}=v_{\rm ml}(t)=L(t)\frac{di_{\rm ml}}{dt}+i_{\rm ml}(t)\frac{dL}{dt}$ should be taken into account. Thus, $U_{\rm ml}=\int \! i_{\rm ml}L \, di_{\rm ml}+\int \! i_{\rm ml}^2 \, dL $. Similar to MC systems, in principle a ML system can be a passive, an active, and even a dissipative element and using the same approach, under some assumptions they behave like passive elements~\citep{Putting_Memory_Into_Circuit_Elements}. There are two types of control process,

	\begin{description}
	\item[I.] $n$th order current-controlled ML systems, $q_{\rm ml}=\int \! i_{\rm ml}(\tau) \, d\tau$,
		\begin{itemize}
		\item $\varphi_{\rm ml}(t)=L(x,i_{\rm ml},t)i_{\rm ml}(t)$
		\item $\dot{x}=f_{\rm icml}(x,i_{\rm ml},t)$
		\end{itemize}
	\item[II.] $n$th order flux-controlled ML systems, $\varrho_{\rm ml}=\int \! \varphi_{\rm ml}(\tau) \, d\tau=\iint \! v_{\rm ml}(\tau) \, d\tau$,
		\begin{itemize}
		\item $i_{\rm ml}(t)=L^{-1}(x,\varphi_{\rm ml},t)\varphi_{\rm ml}(t)$
		\item $\dot{x}=f_{\rm fcml}(x,\varphi_{\rm ml},t)$ .
		\end{itemize}
	\end{description}

\end{description}
	
	\citet{SPICE_modeling_of_memristive_memcapacitative_and_meminductive} introduced a generic SPICE model for mem-devices. Their memristor model was discussed before in this section. Unfortunately, there are no simulation results available in~\citet{SPICE_modeling_of_memristive_memcapacitative_and_meminductive} for MC and ML systems.
	
	It is worth noting that there is no equivalent mechanical element for the memristor. In~\citet{The_missing_mechanical_circuit_element} a new mechanical suspension component, named a {\it J-damper}, has been studied. This new mechanical component has been introduced and tested in Formula One Racing, delivering significant performance gains in handling and grip~\citep{The_missing_mechanical_circuit_element}. In that paper, the authors attempt to show that the {\it J-damper}, which was invented and used by the McLaren team, is in fact an {\it inerter}~\citep{Synthesis_of_mechanical_networks_The_inerter}. The {\it inerter} is a one-port mechanical device where the equal and opposite applied force at the terminals is proportional to the relative acceleration between them~\citep{Synthesis_of_mechanical_networks_The_inerter}. Despite the fact that the {\it missing mechanical circuit element} has been chosen because of a ``spy scandal'' in the 2007 Formula One race~\citep{Formula1}, it may be possible that a ``real'' missing mechanical equivalent to a memristor may someday be found, as its mechanical model is described by~\citet{Oster_A_New_Bond_Graph_Elementr} and~\citet{c24}.
	
	
\section{SPICE Macro-Model of memristor}
\label{spice}

	Basically, there are three different ways available to model the electrical characteristics of the memristors. SPICE macro-models, hardware description language (HDL), and C programming. The first, SPICE macro-models, approach is more appropriate since it is more readable for most of the readers and available in all SPICE versions. There is also another reason for choosing SPICE modeling approach. Regardless of common convergence problems in SPICE modeling, we believe it is more appropriate way to describe real device operation. Moreover, using the model as a sub-circuit can highly guarantee a reasonably high flexibility and scalability features.
	
\begin{figure}[thpb]
		\centering
		\setlength{\unitlength}{1cm}
			\begin{picture}(6.0,3.4)(1.5,0)
				\put(1,0){\includegraphics[scale=.4]{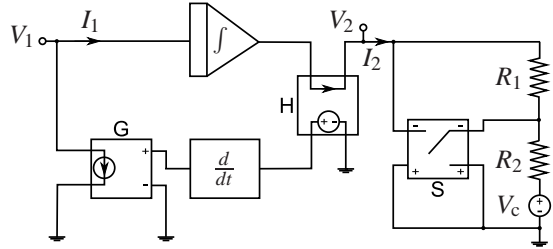}}
				\put(0.8,2.8){\makebox(0,0){{$V_1$}}}
				\put(1.7,3.0){\makebox(0,0){{$I_1$}}}
				\put(5.0,3.0){\makebox(0,0){{$V_2$}}}
				\put(5.4,2.5){\makebox(0,0){{$I_2$}}}
				\put(3.4,2.75){\makebox(0,0){{$\int$}}}				
				\put(3.45,1.05){\makebox(0,0){{$\frac{d}{dt}$}}}
				
				\put(7.2,1.15){\makebox(0,0){{$R_2$}}}
				\put(7.2,2.25){\makebox(0,0){{$R_1$}}}
				\put(7.2,0.55){\makebox(0,0){{$V_{\rm c}$}}}						
			\end{picture}		
		\caption{The SPICE macro-model of memristor. Here G, H and S are a Voltage-Controlled Current Source (VCCS), a Current-Controlled Voltage Source (CCVS), and a Switch ($V_{\rm ON}=-1.9~{\rm V}$ and $V_{\rm OFF}=-2~{\rm V}$), respectively. $R_1 = R_2 = 1~{\rm k}\Omega$ and $V_{\rm c} = -2~{\rm V}$. The M-R mutator consists of an integrator, a Current-Controlled Voltage Source (CCVS) ``H'', a differentiator and a Voltage-Controlled Current Source (VCCS) ``G''. The nonlinear resistor is also realised with resistors $R_1$, $R_2$, and a switch. Therefore, the branch resistance is $1~{\rm k} \Omega$ for $V < 2~{\rm Volt}$ and $2~{\rm k} \Omega$ for $V \geq 2~{\rm Volt}$.}
		\label{fig10}
	\end{figure}

A memristor can be realised by connecting an appropriate nonlinear resistor, inductor, or capacitor across port 2 of an M-R mutator, an M-L mutator, or an M-C mutator, respectively~\footnote{For further detail about the mutator the reader is referred to~\citet{c33}.}~\citep{c16}. These mutators are nonlinear circuit elements that may be described by a SPICE macro-model (i.e. an analog behavioural model of SPICE). The macro circuit model realisation of a type-1 M-R mutator based on the first realisation of the memristor \citep{c16} is shown in Fig.~\ref{fig10}.
	
	In this model the M-R mutator consists of an integrator, a Current-Controlled Voltage Source (CCVS) ``H'', a differentiator and a Voltage-Controlled Current Source (VCCS) ``G''. The nonlinear resistor is also realised with resistors $R_1$, $R_2$, and a switch. Therefore, the branch resistance is $1~{\rm k} \Omega$ for $V < 2~{\rm Volt}$ and $2~{\rm k} \Omega$ for $V \geq 2~{\rm Volt}$. The input voltage of port 1, $V_1$, is integrated and connected to port 2 and the nonlinear resistor current, $I_2$, is sensed with the CCVS ``H'' and differentiated and converted into current with the VCCS ``G''.

\begin{figure}[thpb]
		\centering
		\includegraphics[scale=.5]{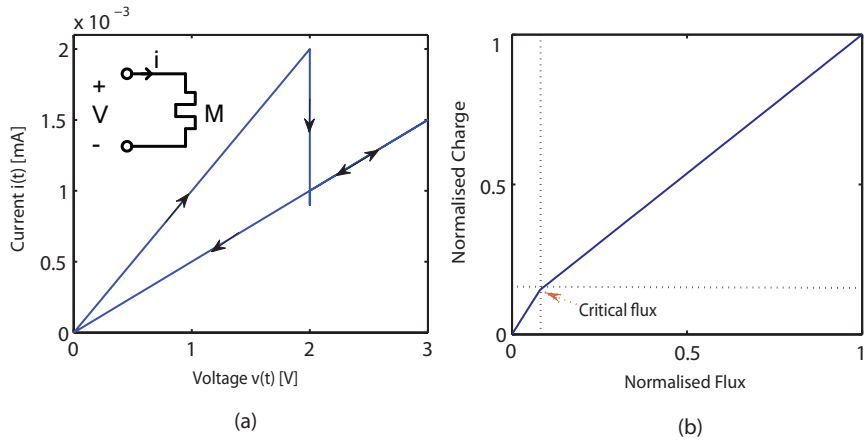}
		\caption{The memristor characteristics. (a) The hysteresis characteristics of the memristor. (b) A monotonically-increasing and piecewise-linear $q$-$\varphi$ function as a basic memristor characteristic, the both axes are normalised to their maximum values. The simulated memristor has a value of $1~{\rm k} \Omega$ when the flux is less than $2~{\rm Wb}$, but it becomes $2~{\rm k} \Omega$ when the flux is equal or higher than $2~{\rm Wb}$. The critical flux ($\varphi_{\rm c}$) can be varied with the turn-on voltage of the switch in the macro-model. The input voltage to the memristor is a ramp with a slope of $\pm 1~{\rm V}/{\rm s}$.}
		\label{figMemCharFig11}
	\end{figure}

\begin{figure}[thpb]
		\centering
		\includegraphics[scale=.5]{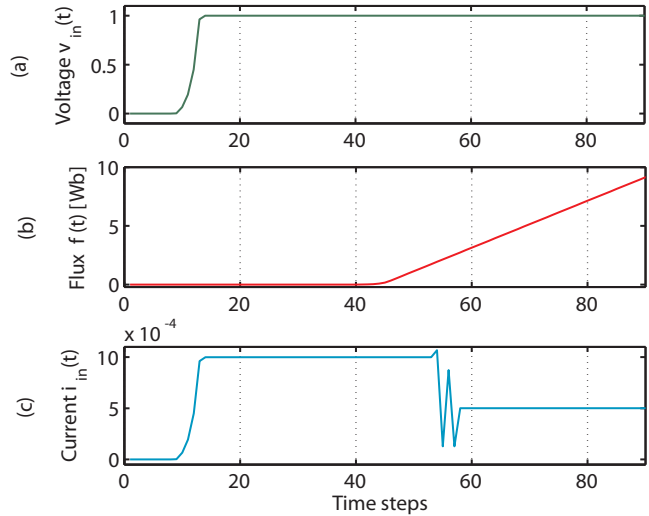}
		\caption{The memristor characteristics when a step input voltage is applied. (a) Voltage curve. (b) Flux linkage curve. (c) Current curve. At the first point the memristance is $1~{\rm k} \Omega$, so the input current is $1~{\rm mA}$. When the flux reaches to $2~{\rm Wb}$ ($1~{\rm V}\times 2~{\rm s}$, $2~{\rm s}=50$ time steps), the memristance is $2~{\rm k} \Omega$ and so the input current is now $0.5~{\rm mA}$ as predicted.}
		\label{fig12}
	\end{figure}

	SPICE simulations with the macro-model of the memristor are shown in Figs.~\ref{figMemCharFig11} and~\ref{fig12}. In this particular simulation, a monotonically-increasing and piecewise-linear $q$-$\varphi$ function is assumed as the memristor characteristic. This function is shown in Fig.~\ref{figMemCharFig11} (b). The simulated memristor has a value of $1~{\rm k} \Omega$ when the flux is less than $2~{\rm Wb}$, but it becomes $2~{\rm k} \Omega$ when the flux is equal or higher than $2~{\rm Wb}$. The critical flux ($\varphi_{\rm c}$) can be varied with the turn-on voltage of the switch in the macro-model. Fig.~\ref{figMemCharFig11} (a) shows the pinched hysteresis characteristic of the memristor. The input voltage to the memristor is a ramp with a slope of $\pm 1~{\rm V}/{\rm s}$. When the input voltage ramps up with a slope of $\pm 1~{\rm V}/{\rm s}$, the memristance is $1~{\rm k} \Omega$ and the slope of the current-voltage characteristics is $1~{\rm mA}/{\rm V}$ before the the flux reaches to the $\varphi_{\rm c}$. But when the flux becomes $2~{\rm Wb}$, the memristance value is changed to $2~{\rm k} \Omega$ and the slope is now $0.5~{\rm mA}/{\rm V}$. After the input voltage reaches to the maximum point, it ramps down and the slope is maintained, because the memristance is still $2~{\rm k} \Omega$. Fig.~\ref{fig12} shows the memristor characteristics when a step input voltage is applied. Initially the memristance is $1~{\rm k} \Omega$, so the input current is $1~{\rm mA}$. When the flux reaches to $2~{\rm Wb}$ ($1~{\rm V}\times 2~{\rm s}$), the memristance is $2~{\rm k} \Omega$ and so the input current is now $0.5~{\rm mA}$ as predicted. The developed macro-model can be used to understand and predict the characteristics of a memristor.

\begin{figure}[thpb]
		\centering
		\includegraphics[scale=0.5]{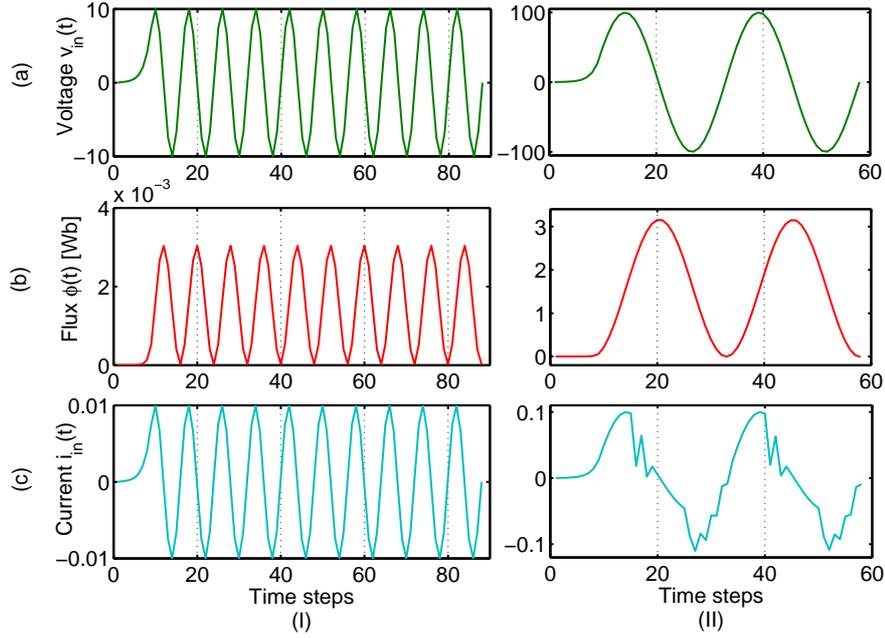}
		\caption{The memristor characteristics when, (I) a $1~{\rm kHz}$ sinusoidal voltage is applied. In this case the flux does not reach to $2~{\rm Wb}$ so $M=M_1=1~{\rm k} \Omega$ and $i=10~{\rm mA}$. (II) When a $10~{\rm Hz}$ sinusoidal voltage is applied. In this case the flux linkage reaches to $2~{\rm Wb}$ within $30~{\rm ms}$, or $15$ time steps. (a) Voltage curve. (b) Flux linkage curve. (c) Current curve.}
		\label{fig12SinBoth}
	\end{figure}

	Now, if a $1~{\rm kHz}$ sinusoidal voltage source is connected across the memristor model, the flux does not reach to $2~{\rm Wb}$ so $M=M_1=1~{\rm k} \Omega$ and $i=10~{\rm mA}$. Fig.~\ref{fig12SinBoth}(I) shows the memristor characteristics when a sinusoidal input voltage is applied. As it is shown in Fig.~\ref{fig12SinBoth}(II), when the voltage source frequency reduces to $10~{\rm Hz}$, the flux linkage reaches to $2~{\rm Wb}$ within $30~{\rm ms}$. Based on this result, there are two levels of memristance, $M=M_1=1~{\rm k} \Omega$ and then it changes to $M_2=2~{\rm k} \Omega$.

\begin{figure}[thpb]
		\centering
		\includegraphics[scale=.5]{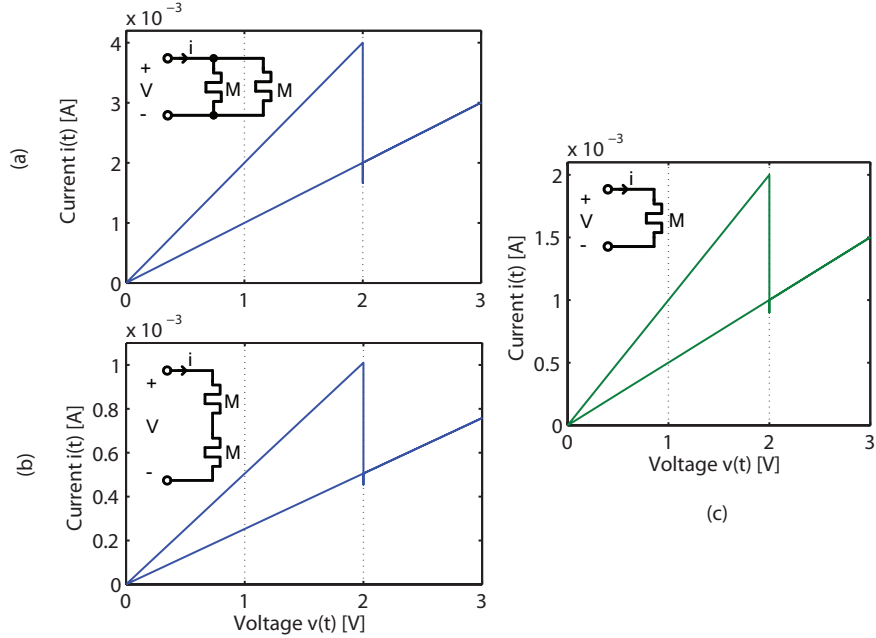}
		\caption{The ${\rm I}$-${\rm V}$ curves for, (a) two memristors in parallel, (b) two memristor in series, and (c) a single memristor. In all of the cases there is no difference between a memristor and equivalent resistor in the network. In other words, two memristors in parallel, with the same characteristics, form a single memristor with $M_{\rm equ}=M_{q}/2$ as memristance, and two memristors in series form a single memristor with $M_{\rm equ}=2M_{q}$ as memristance.}
		\label{figMemristorParalSerial}
	\end{figure}

	Another interesting study is needed to verify that the model is working properly when there is a parallel, series, RM (Resistor-Memristor), LM (Inductor-Memristor), or CM (Capacitor-Memristor) network. First of all, let us assume that there are two memristors with the same characteristic as shown in Fig.~\ref{figMemCharFig11}. Analysing series and parallel configurations of these memristors are demonstrated in Figs.~\ref{figMemristorParalSerial}(b) and \ref{figMemristorParalSerial}(a), respectively. In both figures, the left ${\rm I}$-${\rm V}$ curve shows a single memristor.

\begin{figure}[thpb]
		\centering
		\includegraphics[scale=.5]{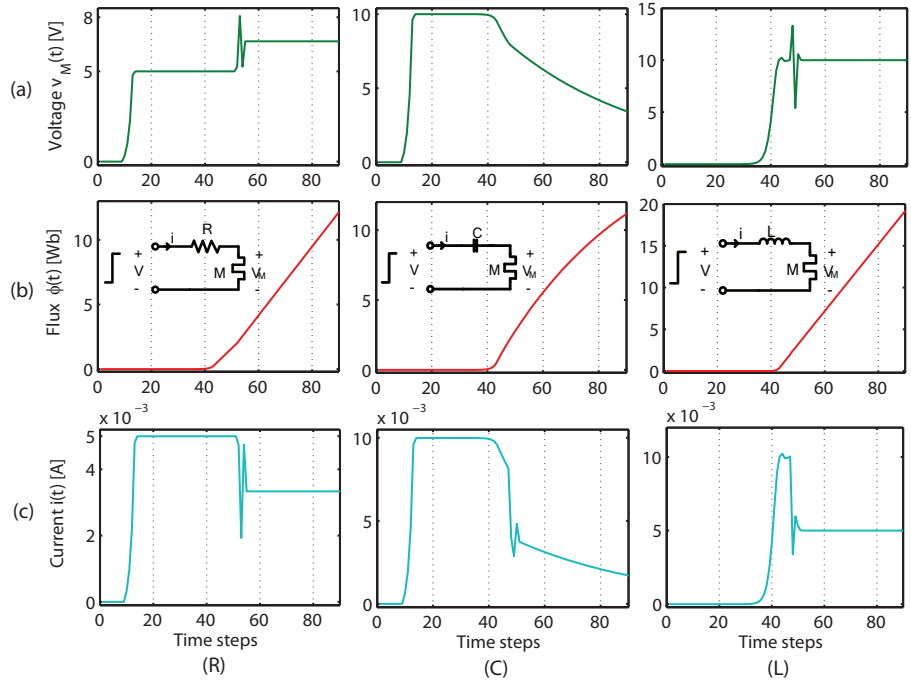}
		\caption{Step voltage response curves for (R) Resistor-Memristor, (C) Capacitor-Memristor, and (L) Inductor-Memristor. (a) Memristor voltage curve, $V_M(t)$. (b) Flux linkage curve, $\varphi(t)$. (c) Current curve, $i_C(t)$. The memristance value switches at time $T_d$, then for $0\leq t\leq T_d$, $\varphi \leq \varphi_{c}$, and $M=M_1=1~{\rm k}\Omega$. Therefore, in the RM circuit we have, $V_M=V\frac{M_1}{R+M_1}$. For $R=1~{\rm k}\Omega$, $V_M=5~{\rm V}$ and then $i_M=5~{\rm mA}$, so $T_d=\frac{\varphi_{c}}{V_M}=0.4~{\rm s}$. Likewise, when $t>T_d$, $\varphi > \varphi_{c}$, we have, $V_M=V\frac{M_2}{R+M_2}=6.7~{\rm V}$ and $i_M=3.3~{\rm mA}$. In the LM circuit we have the same situation, so for $0\leq t\leq T_d$, $\varphi \leq \varphi_{c}$, $V_M=V=10~{\rm V}$, $i_M=10~{\rm mA}$ ($R=1~{\rm k}\Omega$), and $T_d=0.2~{\rm s}$. For $t> T_d$, memristor current is $i_M=5~{\rm mA}$.}
		\label{figRMCMLM}
	\end{figure}

	The simulation results show that the series and parallel configurations of memristors are the same as their resistor counterparts. It means the equivalent memristances, $M_{\rm eq}$, of a two memristor in series and parallel are $M_{\rm eq}=2M$ and $M_{\rm eq}=M/2$, respectively, where $M$ is memristance of a single memristor. The second step is RM, LM, and CM networks. In these cases a $10~{\rm V}$ step input voltage has been applied to circuits. As mentioned before for a single memristor based on the proposed model, while the flux linkage is less than or equal to the critical flux, $\varphi \leq \varphi_{c}$, $M=M_1=1~{\rm k}\Omega$ and when the flux is more than the critical flux value, $\varphi > \varphi_{c}$, $M=M_2=2~{\rm k}\Omega$. Recall that the critical flux value based on the $q-\varphi$ curve is $\varphi_{c}=2~{\rm Wb}$. Figs.~\ref{figRMCMLM}(R), \ref{figRMCMLM}(C), and \ref{figRMCMLM}(L) illustrate RM, CM, and LM circuits and their response to the input step voltage, respectively.

If we assume that the memristance value switches at time $T_d$, then for $0\leq t\leq T_d$, $\varphi \leq \varphi_{c}$, and $M=M_1=1~{\rm k}\Omega$. Therefore, in the RM circuit we have, $V_M=V\frac{M_1}{R+M_1}$. For $R=1~{\rm k}\Omega$, $V_M=5~{\rm V}$ and then $i_M=5~{\rm mA}$, so $T_d=\frac{\varphi_{c}}{V_M}=0.4~{\rm s}$. Likewise, when $t>T_d$, $\varphi > \varphi_{c}$, we have, $V_M=V\frac{M_2}{R+M_2}=6.7~{\rm V}$ and $i_M=3.3~{\rm mA}$. Both cases have been verified by the simulation results.

In the LM circuit we have the same situation, so for $0\leq t\leq T_d$, $\varphi \leq \varphi_{c}$, $V_M=V=10~{\rm V}$, $i_M=10~{\rm mA}$ ($R=1~{\rm k}\Omega$), and $T_d=0.2~{\rm s}$. For $t> T_d$, memristor current is $i_M=5~{\rm mA}$. Memristor's current changing is clearly shown in Fig.~\ref{figRMCMLM}(L). The CM circuit simulation also verifies a change in time constant from $M_1C$ to $M_2C$.

\begin{figure}[thpb]
		\centering
		\includegraphics[scale=.5]{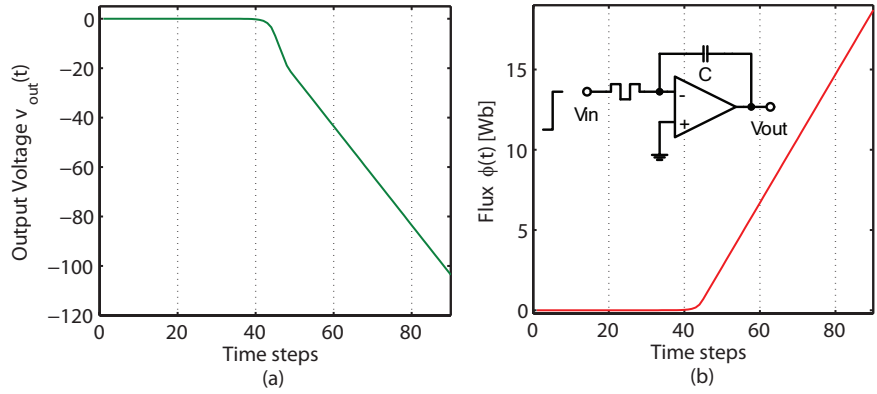}
		\caption{Memristor-op-amp integrator circuit and its response to the input step voltage. (a) Output voltage curve, $V_{\rm out}(t)$. (b) Flux linkage curve, $\varphi(t)$. If $C=100~\mu{\rm C}$ then we have $0.1~{\rm s}$ and $0.2~{\rm s}$ as the time constant of the circuit at $t\leq T_d$ and $t> T_d$, respectively.}
		\label{figOPAMP}
	\end{figure}

As another circuit example of using the new memristor model, an op-amp integrator has been chosen. The model of an op-amp and circuit configuration is shown in Fig.~\ref{figOPAMP}. If $C=100~\mu{\rm C}$ then we have $0.1~{\rm s}$ and $0.2~{\rm s}$ as the time constant of the circuit at $t\leq T_d$ and $t> T_d$, respectively.

It is worth mentioning that, recently, a few simple SPICE macro-models have been proposed by~\citet{On_SPICE_macromodelling_of_Ti},~\citet{SPICE_Model_of_memristor_with_nonlinear_dopant_drift} and~\citet{ICCCAS09_Approximated_SPICE_Model_for_Memristor} but none of them consider the model response with different circuit elements types, which is an important step to verify the model correctness.


\section{Interpreting memristor in Electromagnetic Theory}
\label{maxwell}
	
In his original paper~\citet{c16} presented an argumentation based on electromagnetic field theory for the existence of the memristor. His motivation was to interpret the memristor in terms of the so-called \textit{quasi-static expansion} of Maxwell's equations. This expansion is usually used to give an explanation to the elements of circuit theory within the electromagnetic field theory.

Chua's argumentation hints that a memristor might exist, it although never proved that this device can in fact be realised physically. In the following we describe how Chua argued for a memristor from a consideration of quasi-static expansion of Maxwell's equations. To consider this expansion, we use Maxwell's equations in their differential form,

\begin{eqnarray}
	\nabla \cdot \mathbf{D} = \rho~, \label{equ:maxwell:gausslaw} \\
	\nabla \cdot \mathbf{B} = 0~, \label{equ:maxwell:gausslawofmag}
\\
	\nabla \times \mathbf{E} = -\frac{\partial \mathbf{B}}{\partial t}~, \label{equ:maxwell:faradlaw}
\\
\nabla \times \mathbf{H} = \mathbf{J} + \frac{\partial \mathbf{D}}{\partial t}~, \label{equ:maxwell:amperelaw}
\end{eqnarray}
where $\mathbf{E}$ is electric field intensity (${\rm V/m}$), $\mathbf{B}$ is magnetic flux density (${\rm Wb/m^2}$), $\mathbf{J}$ is electric current density (${\rm A/m^2}$), $\rho$ is electric charge density (${\rm C/m^3}$), $\mathbf{H}$ and $\mathbf{D}$ are magnetic field intensity (${\rm A/m}$) and electric flux density (${\rm C/m^2}$), $\nabla \cdot$ and $\nabla \times$ are divergence and curl operators.

The idea of a quasi-static expansion involves using a process of successive approximations for time-varying fields. The process allows us to study electric circuits in which time variations of electromagnetic field are slow, which is the case for electric circuits.

Consider an entire family of electromagnetic fields for which the time rate of change is variable. The family of fields can be described by a time-rate parameter $\alpha $ which is time rate of change of charge density $\rho $. We can express Maxwell's equations in terms of the {\it family time}, $\tau =\alpha t$, and the time derivative of $\mathbf{B}$ can be written as

\begin{equation}
\frac{\partial \mathbf{B}}{\partial t}=\frac{\partial \mathbf{B}}{\partial
\tau }\frac{d\tau }{dt}=\alpha \frac{\partial \mathbf{B}}{\partial \tau }.
\end{equation}

Other time derivatives can be expressed similarly. In terms of the family time, Maxwell's equations take the form

\begin{equation}
\mathbf{\nabla }\times \mathbf{E=-}\alpha \frac{\partial \mathbf{B}}{\partial \tau }, \text{ \ \ }\mathbf{\nabla }\times \mathbf{H=J+} \alpha \frac{\partial \mathbf{D}}{\partial \tau },  \label{MEqs-1}
\end{equation}
which allow us to consider different values of the family time $\tau $, corresponding to different time scales of excitation. Note that in Eqs.~\ref{MEqs-1} $\mathbf{E}$, $\mathbf{H}$, $\mathbf{D}$, $\mathbf{J}$, and $\mathbf{B}$ are also functions of $\alpha $ and $\tau $, along with the position $(x,y,z)$.This allows us~\citep{c32} to express, for example, $\mathbf{E}(x,y,z,\alpha ,\tau )$ as power series in $\alpha $:

\begin{equation}
\mathbf{E}(x,y,z,\alpha ,\tau )=\mathbf{E}_{0}(x,y,z,\tau )+\alpha \mathbf{E}_{1}(x,y,z,\tau )+\alpha^{2}\mathbf{E}_{2}(x,y,z,\tau )+...,
\label{PowerSeries}
\end{equation}
where the zero and first orders are

\begin{eqnarray}
\mathbf{E}_{0}(x,y,z,\tau ) &=&[\mathbf{E}(x,y,z,\alpha ,\tau )]_{\alpha =0} = - \nabla \Phi_0,
\nonumber \\
\mathbf{E}_{1}(x,y,z,\tau ) &=&\left[ \frac{\partial \mathbf{E}(x,y,z,\alpha,\tau )}{\partial \alpha }\right] _{\alpha =0} = - \frac{\partial \mathbf{A}_0}{\partial \tau}.  \nonumber \\
&&...  \label{PS}
\end{eqnarray}

Along with these, a similar series of expressions for $\mathbf{B}$, $\mathbf{H}$, $\mathbf{J}$, and $\mathbf{D}$ are obtained and can be inserted into Eqs.~\ref{MEqs-1}, with the assumption that every term in these series is
differentiable with respect to $x,$ $y,$ $z,$ and $\tau $. This assumption permits us to write, for example,

\begin{equation}
\mathbf{\nabla }\times \mathbf{E=\nabla }\times \mathbf{E}_{0}+\alpha (\mathbf{\nabla }\times \mathbf{E}_{1})+\alpha ^{2}(\mathbf{\nabla }\times \mathbf{E}_{2})+...,
\end{equation}
and, when all terms are collected together on one side, this makes Eqs.~\ref{MEqs-1} to take the form of a power series in $\alpha $ that is equated to zero. For example, the first equation in Eq.~\ref{MEqs-1} becomes

\begin{equation}
\mathbf{\nabla }\times \mathbf{E}_{0}+\alpha (\mathbf{
\nabla }\times \mathbf{E}_{1}+\frac{\partial \mathbf{B}_{0}}{
\partial \tau })+\alpha ^{2}(\mathbf{\nabla }\times \mathbf{E}_{2}+\frac{\partial \mathbf{B}_{1}}{\partial \tau })+...=0,  \label{PS-1}
\end{equation}
which must hold for all $\alpha $. This can be true if the coefficients of
all powers of $\alpha $ are separately equal to zero. The same applies to
the second equation in Eqs.~\ref{MEqs-1} and one then obtains the so-called $n$
th-order Maxwell's equations, where $n=0,1,2,...$ For instance, the \textit{zero-order} Maxwell's equations are

\begin{eqnarray}
\mathbf{\nabla }\times \mathbf{E}_{0} &=&0,  \label{zero-1st} \\
\mathbf{\nabla }\times \mathbf{H}_{0} &=&\mathbf{J}_{0},
\label{zero-2nd}
\end{eqnarray}%

and the \textit{first-order} Maxwell's equations are

\begin{eqnarray}
\mathbf{\nabla }\times \mathbf{E}_{1} &=&-\frac{\partial \mathbf{B}%
_{0}}{\partial \tau },  \label{first-1st} \\
\mathbf{\nabla }\times \mathbf{H}_{1} &=&\mathbf{J}_{1}+\frac{%
\partial \mathbf{D}_{0}}{\partial \tau }.  \label{first-2nd}
\end{eqnarray}

The quasi-static fields are obtained from only the first two terms of the
power series Eq.~\ref{PS-1}, while ignoring all the remaining terms and by
taking $\alpha =1$. In this case we can approximate $\mathbf{E\approx E}_{0}+%
\mathbf{E}_{1},$ $\mathbf{D\approx D}_{0}+\mathbf{D}_{1},$ $\mathbf{H\approx
H}_{0}+\mathbf{H}_{1},$ $\mathbf{B\approx B}_{0}+\mathbf{B}_{1},$ and $%
\mathbf{J\approx J}_{0}+\mathbf{J}_{1}$.

Circuit theory, along with many other electromagnetic systems, can be explained by the zero-order and first-order Maxwell's equations for which one obtains \textit{quasi-static fields} as the solutions. The three classical circuit elements \textit{resistor}, \textit{inductor}, and \textit{capacitor} can then be explained as electromagnetic systems whose quasi-static solutions correspond to certain combinations of the zero-order and the first-order solutions of Eqs.~\ref{zero-1st}-\ref{first-2nd}.

However, in this quasi-static explanation of circuit elements, an interesting possibility was unfortunately dismissed~\citep{c32} as it was thought not to have any correspondence with an imaginable situation in circuit
theory. This is the case when both the first-order electric and the first-order magnetic fields are \textit{not} negligible. Chua argued that it is precisely this possibility that provides a hint towards the existence a fourth basic circuit device.

Chua's argumentation goes as follows. Assume there exists a two-terminal device in which $\mathbf{D}_{1}$ is related to $\mathbf{B}_{1}$, where these quantities are evaluated instantaneously. If this is the case then this device has the following two properties:

\begin{enumerate}
    \item Zero-order electric and magnetic fields are negligible when compared to the first-order fields i.e. $\mathbf{E}_{0},$ $\mathbf{D}_{0},$ $\mathbf{B}_{0},$ and $\mathbf{J}_{0}$ can be ignored.
    \item The device is made from \textit{non-linear} material for which the first-order fields become related.
\end{enumerate}

Assume that the relationships between the first-order fields are expressed as

\begin{eqnarray}
\mathbf{J}_{1} &=&\mathcal{J}(\mathbf{E}_{1}),  \label{non-linear-1} \\
\mathbf{B}_{1} &=&\mathcal{B}(\mathbf{H}_{1}),  \label{non-linear-2} \\
\mathbf{D}_{1} &=&\mathcal{D}(\mathbf{E}_{1}),  \label{non-linear-3}
\end{eqnarray}%
where $\mathcal{J},$ $\mathcal{B},$ and $\mathcal{D}$ are one-to-one continuous functions defined over space coordinates only. Combining Eq.~\ref{first-2nd}, in which we have now $\mathbf{D}_{0}\approx 0$, with Eq.~\ref{non-linear-1} gives

\begin{equation}
\mathbf{\nabla }\times \mathbf{H}_{1}=\mathcal{J}(\mathbf{E}_{1}).
\label{HE-first}
\end{equation}
As the curl operator does not involve time derivatives, and $\mathcal{J}$ is
defined over space coordinates, Eq.~\ref{HE-first} says that the
first-order fields $\mathbf{H}_{1}$ and $\mathbf{E}_{1}$ are related. This
relation can be expressed by assuming a function $\mathcal{F}$ and we can
write

\begin{equation}
\mathbf{E}_{1}=\mathcal{F(}\mathbf{H}_{1}).  \label{HE-2nd}
\end{equation}
Now, Eq.~\ref{non-linear-3} can be re-expressed by using Eq.~\ref{HE-2nd} as

\begin{equation}
\mathbf{D}_{1}=\mathcal{D\circ F(}\mathbf{H}_{1}),  \label{DH}
\end{equation}
where $\circ$ operator is the composition of two (or more) functions. Also, as $\mathcal{B}$ is a one-to-one continuous function, Eq.~\ref{non-linear-2} can be re-expressed as

\begin{equation}
\mathbf{H}_{1}=\mathcal{B}^{-1}(\mathbf{B}_{1}).  \label{HB}
\end{equation}%
Inserting from Eq.~\ref{HB} into Eq.~\ref{DH} then gives

\begin{equation}
\mathbf{D}_{1}=\mathcal{D\circ F\circ }\left[ \mathcal{B}^{-1}(\mathbf{B}%
_{1})\right] \equiv \mathcal{G(}\mathbf{B}_{1}).  \label{DB}
\end{equation}

Eq.~\ref{DB} predicts that an instantaneous relationship can be
established between $\mathbf{D}_{1}$ and $\mathbf{B}_{1}$ that is realisable in a memristor. This completes Chua's argumentation using Maxwell's equations for a quasi-static representation of the electromagnetic field quantities of a memristor.


\section{Conclusion} \label{conclusion} In this paper, we surveyed key aspects of the memristor as a promising nano-device. We also introduced a behavioural and SPICE macro-model for the memristor and reviewed Chua's argumentation for the memristor by performing a quasi-static expansion of Maxwell's equations. The SPICE macro-model has been simulated in PSpice and shows agreement with the actual memristor presented in~\citet{c17}. The model shows expected results in combination with a resistor, capacitor, or inductor. A new op-amp based memristor is also presented and tested.
	
 Nanoelectronics not only deals with the nanometer scale, materials, and devices but implies a revolutionary change even in computing algorithms. The Von-Neumann architecture is the base architecture of all current computer systems. This architecture will need revision for carrying out computation with nano-devices and materials. There are many of different components, such as processors, memories, drivers, actuators and so on, but they are poor at mimicking the human brain. Therefore, for the next generation of computing, choosing a suitable architecture is the first step and requires deep understanding of relevant nano-device capabilities. Obviously, different capabilities might create many opportunities as well as challenges. At present, industry has pushed nanoelectronics research for highest possible compatibility with current devices and fabrication processes. However, the memristor motivates future work in nanoelectronics and nano-computing based on its capabilities.

	In this paper we addressed some possible research gaps in the area of the memristor and demonstrated that further device and circuit modelling are urgently needed. The current approach to device modelling is to introduce a physical circuit model with a number of curve fitting parameters. However, such an approach has the limitation of requiring a large number of parameters. Using a non-linear drift model results in more accurate simulation at the cost of a much more complicated set of mathematical equations. Initially behavioural modelling can be utilised, nonetheless a greater modelling effort is needed to accommodate both the defect and process variation issues. An interesting follow up would be the development of mapping models based on the memristor to neuronmorphic systems that deal with architectural level challenges, such as defect-tolerance and integration into current integrated circuit technologies.

\section{Acknowledgements} \label{ack} We would like to thank Leon Chua, at UC Berkeley, for useful discussion and correspondence. This work was partially supported by grant No. R33-2008-000-1040-0 from the World Class University (WCU) project of MEST and KOSEF through CBNU is gratefully acknowledged. 	

\bibliographystyle{abbrvnat}
\bibliography{bib_memristor}

\end{document}